\documentclass[preprint,a4paper,12pt]{elsarticle}
\pdfoutput=1
\usepackage[charter,greekuppercase=italicized]{mathdesign}
\usepackage{hyperref}

\journal{Journal of Magnetism and Magnetic Materials}

\begin{document}

\begin{frontmatter}

\title{Structure, magnetic and thermodynamic properties of heterometallic ludwigites: Cu$_2$GaBO$_5$ and Cu$_2$AlBO$_5$}

\author[1]{R.~M.~Eremina\corref{cor1}}
\cortext[cor1]{REremina@yandex.ru}

\author[1]{T.~P.~Gavrilova}

\author[2]{E.~M.~Moshkina}
\author[3]{I.~F.~Gilmutdinov}
\author[3]{R.~G.~Batulin}

\author[4]{V.~V.~Gurzhiy}

\author[7,8]{V.~Grinenko}

\author[8]{D.~S.~Inosov}

\address[1]{Zavoisky Physical-Technical Institute, FRC Kazan Scientific Center of RAS, Sibirsky tract, 10/7, Kazan, 420029, Russia}
\address[2]{Kirensky Institute of Physics, Federal Research Center KSC SB RAS, 660036 Krasnoyarsk, Russia}
\address[3]{Institute of Physics, Kazan Federal University, 420008, Kazan, Russia}
\address[4]{Department of Crystallography, Institute of Earth Sciences, St. Petersburg State University, University Emb. 7/9, 199034, St. Petersburg, Russia}
\address[7]{Leibniz Institute for Solid State and Materials Research, IFW Dresden, 01069, Dresden, Germany}
\address[8]{Institute for Solid State and Materials Physics, TU Dresden, 01069, Dresden, Germany}

\date{\today}
             
\begin{abstract}
We present an extensive study of the structural, magnetic and thermodynamic properties of high-quality monocrystals of the two heterometallic oxyborates from the ludwigite family: Cu$_2$GaBO$_5$ and Cu$_2$AlBO$_5$
in the temperature range above 2\,K. The distinctive feature of the investigated structures is the selective distribution of Cu and Ga/Al cations. The unit cell of Cu$_2$GaBO$_5$ and Cu$_2$AlBO$_5$ contains four nonequivalent crystallographic sites of metal ions. Two sites in the structure from four nonequivalent crystallographic sites of metal ions of Cu$_2$GaBO$_5$ are fully occupied by Cu atoms which form the quasi one-dimensional chains along the $a$-axis. For Cu$_2$AlBO$_5$ all sites are partially  occupied by Al and Cu atoms. The joint analysis of low-temperature data on magnetic susceptibility and magnetic contribution to the specific heat showed that Cu$_2$AlBO$_5$ and Cu$_2$GaBO$_5$ exhibit an antiferromagnetic transition at $T_{\rm N}\approx$\,3 and 4\,K, respectively. The magnetic contributions to the specific heat for both compounds were obtained after subtracting the phonon contribution. It is shown that the external magnetic field above 2.5\,T leads to a broadening of the magnetic phase transition indicating suppression of the long-range antiferromagnetic order.
\end{abstract}


\end{frontmatter}

\section{Introduction}

Cu$_2$GaBO$_5$ and Cu$_2$AlBO$_5$ oxyborates belong to the ludwigite family with the general formula $M^{2+}_{2}M'^{3+}$BO$_5$, where $M$ and $M'$ are divalent and trivalent metal ions, respectively. During the last twenty years a lot of works were devoted to the investigations of bimagnetic ludwigites Cu$_2$MnBO$_5$~\cite{Sofronova_2016, Moshkina_2017,Moshkina_2018}, Mn$_{3-x}$Ni$_{x}$BO$_5$~\cite{Bezmaternyk_2014},
Cu$_2$FeBO$_5$~\cite{Continentino_1999,Nazarenko_2018}, Ni$_2$FeBO$_5$~\cite{Fernandes_1998,Freitas_2009}, Co$_2$FeBO$_5$~\cite{Freitas_2009} and others.
Usually the investigations of ludwigites start from the detailed sample characterization, because the final sample composition can differ
from the composition of the corresponding mixture of the starting components. In addition to the structural data, the magnetization~\cite{Moshkina_2017,Moshkina_2018,Bezmaternyk_2014,Continentino_1999,Fernandes_1998,Freitas_2009},
specific heat~\cite{Moshkina_2017,Freitas_2009}, neutron powder diffraction~\cite{Moshkina_2017}, M\"{o}ssbauer spectroscopy~\cite{Continentino_1999,Fernandes_1998} measurements, and the calculations of the exchange integrals in frameworks of the indirect coupling model~\cite{Sofronova_2016,Nazarenko_2018} are presented in the literature.

Magnetic properties of oxyborates with the ludwigite structure are usually related with the presence of zigzag walls in their crystal structure formed by metal ions of different valency and also the presence of up to twelve magnetic ions in the unit cell, which occupy four nonequivalent positions. Usually the copper-containing ludwigites are characterized by the antiferromagnetic or ferrimagnetic ordering with low value of the uncompensated magnetic moment and low temperature of magnetic ordering.

Partial substitution of copper ions with Co$^{2+}$ cations and occupation of trivalent positions with Al$^{3+}$ cations leads to a significant anisotropy of the magnetic properties in CuCoAlBO$_5$~\cite{Petrakovskii_2009_CuCoAlBO5}. Authors suggested that such a difference is due to the influence of the strong spin-orbit coupling of Co$^{2+}$ ions, which leads to the canting of the magnetic moments on neighboring sublattices and causes a weak spontaneous magnetic moment~\cite{Petrakovskii_2009_CuCoAlBO5}. Unlike other Cu-containing oxyborates, Co$_{2.88}$Cu$_{0.12}$BO$_5$ is the highly anisotropic hard ferrimagnet with a large uncompensated moment~\cite{Ivanova_2013}. However, the comparison of magnetic properties of Co$_{2.88}$Cu$_{0.12}$BO$_5$ ludwigite with homometallic Co$_3$BO$_5$ showed that the replacement of cobalt ions with copper does not affect the magnetic properties of the sample: a slight decrease in the macroscopic magnetic moment and invariability of the ferrimagnetic ordering temperature ($T_{\rm N}=43$~K) were observed~\cite{Ivanova_2013}.
The Cu$_{3-x}$Mn$_{x}$BO$_5$ ($x$\,=\,2) ludwigite is characterized by the ferrimagnetic ordering below $T_{\rm N}$\,=\,92\,K demonstrating a possible increase in the macroscopic magnetic moment and the  magnetic ordering temperature in ludwigites~\cite{Bezmaternykh_2015}. A completely different picture of phase transitions is observed in
Cu$_2$FeBO$_5$ ludwigite, where the phase transition of the iron subsystem from the paramagnetic to the spin glass state was observed at \emph{T}\,=\,63\,K, the Cu$^{2+}$ subsystem passes into a magnetically ordered state at
$T_{\rm N1}$\,=\,38\,K, and only below $T_{\rm N1}$\,=\,20\,K the sample is fully ordered~\cite{Petrakovskii_2009}.

Magnetic measurements and the analysis of exchange interactions in Cu$_2$FeBO$_5$ and Cu$_2$GaBO$_5$ showed that these compounds are antiferromagnetic (AFM) with N\'{e}el temperatures of 32 and 3\,K, respectively~\cite{Continentino_1999,Petrakovskii_2009}. The authors concluded that the magnetic properties of this type of compounds are substantially dependent on the
degree of cation distribution over crystallographic positions. As follows from Ref.~\cite{Petrakovskii_2009}, Cu$_2$GaBO$_5$ is a low-dimensional magnetic system, for which magnetic transition to an attiferromagnetically ordered state was observed in the temperature dependence of magnetic susceptibility. A definitive answer can only be given by studing the temperature dependence of magnetic susceptibility on alternating current
(AC)~\cite{Balanda_2013}, which has not yet been carried out for the Cu$_2$GaBO$_5$ single crystal.
The details of the synthesis process of Cu$_2$AlBO$_5$ ludwigite were previously reported~\cite{Hriljac_1990}, but the magnetic properties of this compound have not been investigated until now. The temperature dependencies of the specific heat were not obtained for  Cu$_2$AlBO$_5$ and Cu$_2$GaBO$_5$.

Here we present the detailed investigations of structural, magnetic and thermodynamic properties of Cu$_2$GaBO$_5$ and Cu$_2$AlBO$_5$  ludwigites,
which contain only one type of magnetic ion Cu$^{2+}$. We suggest that these investigations will help in understanding the type of the magnetic ordering in the homomagnetic heterometallic ludwigites.

\section{Experimental methods and results}

\subsection{Chemical composition}

Small fragments of Cu$_2$GaBO$_5$ (\textbf{1}) and Cu$_2$AlBO$_5$ (\textbf{2}) were crushed, pelletized, and carbon coated. The chemical compositions of the samples were determined using a Hitachi TM 3000 scanning electron microscope equipped with an EDX spectrometer. Analytical calculations \textbf{1}: Atomic ratio from structural data Cu 1.96, Ga 1.04; found by EDX: Cu 1.95, Ga 1.05. Analytical calculations \textbf{2}: Atomic ratio from structural data Cu 1.82, Al 1.18; found by EDX: Cu 1.84, Al 1.16.

In spite of the real atomic ratio Cu:Ga (or Cu:Al) is not 2:1 in the investigated samples, in this work we use the ideal formula Cu$_2$GaBO$_5$ and Cu$_2$AlBO$_5$ instead Cu$_{2.05}$Ga$_{0.95}$BO$_5$ and Cu$_{1.81}$Al$_{1.19}$BO$_5$.

\begin{table}[t!]
\caption{\label{Table_lat_param} Crystallographic data for Cu$_2$GaBO$_5$ (\textbf{1}) and Cu$_2$AlBO$_5$ (\textbf{2}).}\smallskip
\centerline{
\begin{tabular}{c | c c c c c c c c c}
\hline \hline
Compound              &&  \textbf{1}                        &&  \textbf{2} &&                 \\ \hline
Formula               && Cu$_{2.05}$Ga$_{0.95}$BO$_5$       &&  Cu$_{1.81}$Al$_{1.19}$BO$_5$  \\
Crystal system        && monoclinic                         && monoclinic                     \\
\emph{a} (\AA)        && 3.1121(1)                          && 3.0624(2)                      \\
\emph{b} (\AA)        && 11.9238(3)                         && 11.7855(6)                     \\
\emph{c} (\AA)        && 9.4708(2)                          && 9.3747(6)                      \\
$\alpha$ ($^{\circ}$) && 90                                 && 90                             \\
$\beta$ ($^{\circ}$)  && 97.865(1)                          && 97.737(5)                      \\
$\gamma$ ($^{\circ}$) && 90                                 && 90                             \\
\emph{V} (\AA$^2$)    && 348.137(16)                        && 335.27(4)                      \\
Molecular weight      && 287.87                             && 238.11                         \\
Space group           && $P2_1/c$                           && $P2_1/c$                       \\
$\mu$ (mm$^{-1}$)     && 19.595                             && 11.742                         \\
Temperature (K)       && 293(2)                             && 293(2)                         \\
$Z$                     && 4                                  && 4                              \\
$D_{\rm calc}$ (g/cm$^3$) && 5.481                              && 4.717                          \\
Crystal size (mm$^3$) && 0.18\,$\times$\,0.14\,$\times$\,0.09             && 0.22\,$\times$\,0.16\,$\times$\,0.10         \\
Diffractometer        && Bruker Smart                       && Rigaku Oxford \\
                      && Apex II                            && Diffraction                    \\
                      &&                                    && Xcalibur Eos         \\
Radiation             && Mo\,$K_{\alpha}$                     && Mo\,$K_{\alpha}$                  \\
Total reflections     && 9729                               && 1711                            \\
Unique reflections    && 1873                               && 761                             \\
Angle range 2$\Theta$ ($^{\circ}$) && 5.53--79.20            && 5.58--55.00                      \\
Reflections with      && 1755            && 694                             \\
$\mid F_{\rm o} \mid \geq 4F$ \\
$R_{\rm int}$             && 0.0394                             && 0.0437                          \\
$R_{\sigma}$          && 0.0264                             && 0.048                           \\
$R_1$ ($\mid F_{\rm o} \mid \geq 4F$) && 0.0181                   && 0.0299                          \\
$wR_2$ ($\mid F_{\rm o} \mid \geq 4F$)&& 0.0394                   && 0.0636                          \\
$R_1$ (all data)      && 0.0209                             && 0.0323                          \\
$wR_2$ (all data)     && 0.0402                             && 0.0660                          \\
$S$                     && 1.114                              && 1.047                           \\
$\rho_{\rm min}$, $\rho_{\rm max}$ (e/\AA$^3$) && $-0.715$, 0.883     && $-0.820$, 0.928                   \\
ICSD                  && 1884474                            && 1884475                         \\ \hline \hline
\end{tabular}\smallskip}
$R_1=\Sigma ||F_{\rm o}|-|F_{\rm c}||$; $wR_2=\{\Sigma[w(F_{\rm o}^2-F_{\rm c}^2)^2|\Sigma[w(F_{\rm o}^2)^2]\}^{1/2}$; $w=1/[\sigma^2(F_{\rm o}^2)+(aP)^2+bP]$; where $P=(F_{\rm o}^2+2F_{\rm c}^2)/3$; $S=\{\Sigma[w(F_{\rm o}^2-F_{\rm c}^2)]/(n-p)\}^{1/2}$, where $n$ is the number of reflections and $p$ is the number of refinement parameters.\\
\end{table}

\subsection{\label{sec:X-ray diffraction}Single-crystal X-ray diffraction study}

\begin{table}[b!]
\caption{\label{Table2} Atomic coordinates, isotropic displacement parameters (\AA$^2$) and site occupancy factors (s.o.f.) for Cu$_2$GaBO$_5$ (\textbf{1}) and Cu$_2$AlBO$_5$ (\textbf{2}).\smallskip}
\begin{tabular}{l l l l l l }
\hline \hline
\multicolumn{6}{c}{\textbf{1}} \\
Atom          & $x$           & $y$           & $z$           & $U_{eq}$    & s.o.f.   \\  \hline
Cu1           & 0.46491(6)  & 0.71961(2)  & 0.50724(2)  & 0.00621(6)  & 1        \\
Cu2           & 0.500000    & 0.500000    & 1.000000    & 0.00497(6)  & 1        \\
Ga3           & 0.000000    & 0.500000    & 0.500000    & 0.00490(8)  & 0.66     \\
Cu3           & 0.000000    & 0.500000    & 0.500000    & 0.00490(8)  & 0.34     \\
Cu4           & 0.92862(5)  & 0.61907(2)  & 0.77187(2)  & 0.00601(6)  & 0.29     \\
Ga4           & 0.92862(5)  & 0.61907(2)  & 0.77187(2)  & 0.00601(6)  & 0.71     \\

B1            & 0.9641(5)   & 0.86426(11) & 0.73488(17) & 0.0051(2)   & 1        \\
O1            & 0.4585(4)   & 0.64426(8)  & 0.89925(12) & 0.00934(19) & 1        \\
O2            & 1.0289(3)   & 0.46141(8)  & 0.84306(11) & 0.00658(16) & 1        \\
O3            & 0.9152(3)   & 0.76255(8)  & 0.66705(11) & 0.00808(18) & 1        \\
O4            & 0.0074(3)   & 0.63394(8)  & 0.38096(12) & 0.00685(17) & 1        \\
O5            & 0.5519(7)   & 0.57364(15) & 0.6018(2)   & 0.0068(3)   & 0.63     \\
O5A           & 0.4148(12)  & 0.5856(3)   & 0.6286(4)   & 0.0068(3)   & 0.37
   \\\hline \hline
\multicolumn{6}{c}{\textbf{2}} \\
Atom          & $x$           & $y$           & $z$           & $U_{eq}$    & s.o.f.    \\  \hline
Cu1           & 0.46022(14) & 0.71962(4)  & 0.50700(5)  & 0.0070(2)   & 0.88      \\
Al1           & 0.46022(14) & 0.71962(4)  & 0.50700(5)  & 0.0070(2)   & 0.12      \\
Cu2           & 0.500000    & 0.500000    & 1.000000    & 0.0058(3)   & 0.86      \\
Al2           & 0.500000    & 0.500000    & 1.000000    & 0.0058(3)   & 0.14      \\
Cu3           & 0.000000    & 0.500000    & 0.500000    & 0.0062(4)   & 0.34      \\
Al3           & 0.000000    & 0.500000    & 0.500000    & 0.0062(4)   & 0.66      \\
Cu4           & 0.9285(2)   & 0.61643(6)  & 0.76921(8)  & 0.0065(3)   & 0.33      \\
Al4           & 0.9285(2)   & 0.61643(6)  & 0.76921(8)  & 0.0065(3)   & 0.67      \\
B1            & 0.9632(14)  & 0.8634(3)   & 0.7353(5)   & 0.0093(9)   & 1         \\
O1            & 0.4583(10)  & 0.6440(2)   & 0.8951(4)   & 0.0196(8)   & 1         \\
O2            & 1.0139(7)   & 0.4617(2)   & 0.8422(3)   & 0.0116(7)   & 1         \\
O3            & 0.9147(8)   & 0.7604(2)   & 0.6692(3)   & 0.0129(7)   & 1         \\
O4            & \hspace{-1ex}--0.0045(8)  & 0.6327(2)   & 0.3830(3)   & 0.0117(7)   & 1         \\
O5            & 0.562(2)    & 0.5695(5)   & 0.6011(7)   & 0.0130(12)  & 0.58      \\
O5A           & 0.392(3)    & 0.5835(7)   & 0.6295(11)  & 0.0130(12)  & 0.42       \\
\hline \hline
\end{tabular}\vspace{-2pt}
\end{table}

\begin{table}[t!]
\caption{\label{Table_interatomic_dist} Selected bond lengths in the crystal structure of Cu$_2$GaBO$_5$ and Cu$_2$AlBO$_5$.\smallskip}
\centerline{
\begin{tabular}{@{}l@{~~~}r@{}l|l@{~~~}r@{}l@{}}
\hline \hline
\multicolumn{3}{c|}{Cu$_2$GaBO$_5$} &  \multicolumn{3}{c}{Cu$_2$AlBO$_5$} \\\hline
Bond                 & \multicolumn{2}{l|}{Bond length (\AA)} &  Bond        & \multicolumn{2}{l}{Bond length (\AA)} \\\hline
Cu1--O1           &   & 1.9172(10)     & Cu1(Al1)--O1    & & 1.919(3)        \\
Cu1--O3           &  & 1.9840(10)     & Cu1(Al1)--O3     & & 1.977(2)        \\
Cu1--O4           &  & 2.4241(11)     & Cu1(Al1)--O4     & & 2.366(3)        \\
Cu1--O4           &  & 2.0087(10)     & Cu1(Al1)--O4     & & 1.995(3)        \\
Cu1--O5           &  & 1.9591(18)     & Cu1(Al1)--O5     & & 1.983(6)        \\
Cu1--O5A          &  & 1.987(3)       & Cu1(Al1)--O5A    & & 2.000(9)        \\
$\langle$Cu1--O$\rangle$     &   & 2.047          & $\langle$Cu1(Al1)--O$\rangle$ & & 2.040   \\ & & & \\
Cu2--O1           & 2$\times$ & 1.9627(10)  & Cu2(Al2)--O1 & 2$\times$  & 1.957(3) \\
Cu2--O2           & 2$\times$ & 1.9935(10)  & Cu2(Al2)--O2 & 2$\times$  & 2.004(3) \\
Cu2--O2           & 2$\times$ & 2.4082(11) & Cu2(Al2)--O2 & 2$\times$  & 2.344(3) \\
$\langle$Cu2--O$\rangle$ &          &  2.122       &            &            & 2.102  \\ & & & \\
Cu3(Ga3)--O4      & 2$\times$ & 1.9569(10) & Cu3(Al2)--O4  & 2$\times$  & 1.909(3)  \\
Cu3(Ga3)--O5      & 2$\times$ & 2.0485(18) & Cu3(Al2)--O5  & 2$\times$  & 1.927(7)  \\
Cu3(Ga3)--O5      & 2$\times$ & 2.003(2)   & Cu3(Al2)--O5  & 2$\times$  & 2.022(6)  \\
Cu3(Ga3)--O5A     & 2$\times$ & 1.939(3)   & Cu3(Al2)--O5A & 2$\times$  & 1.869(10) \\
$\langle$Cu3(Ga3)--O$\rangle$  &         & 1.986      & $\langle$Cu3(Al3)--O$\rangle$ &      & 1.932 \\ & & & \\
Cu4(Ga4)--O1      &  & 2.0419(13)    &  Cu4(Al4)--O1     & & 2.007(3)        \\
Cu4(Ga4)--O1      &  & 1.9287(11)    &  Cu4(Al4)--O1     & & 1.901(3)        \\
Cu4(Ga4)--O2      &  & 2.0071(10)    &  Cu4(Al4)--O2     & & 1.953(3)        \\
Cu4(Ga4)--O3      &  & 1.9756(10)    &  Cu4(Al4)--O3     & & 1.936(3)        \\
Cu4(Ga4)--O5      &  & 1.9357(18)    &  Cu4(Al4)--O5     & & 1.891(7)        \\
Cu4(Ga4)--O5A     &  & 2.203(4)      &  Cu4(Al4)--O5A    & & 1.996(10)       \\
Cu4(Ga4)--O5A     &  & 1.991(3)      &  Cu4(Al4)--O5A    & & 2.095(10)       \\
$\langle$Cu4(Ga4)--O$\rangle$ &   & 2.012         & $\langle$Cu4(Al4)--O$\rangle$ & & 1.968 \\ & & & \\
B1--O2            &  & 1.3754(17)        & B1--O2            & & 1.374(5) \\
B1--O3            &  & 1.3711(18)        & B1--O3            & & 1.362(5) \\
B1--O4            &  & 1.3717(19)        & B1--O4            & & 1.376(5) \\
$\langle$B1--O$\rangle$ &  & 1.373             & $\langle$B1--O$\rangle$    & & 1.371    \\
\hline \hline
\end{tabular}}
\end{table}

Crystal structures of \textbf{1} and \textbf{2} were determined by the means of single-crystal X-ray diffraction analysis. Crystals were selected under an optical microscope, encased in oil-based cryoprotectant, and fixed on micro mounts. Diffraction data for \textbf{1} were collected at 293~K using a Bruker SMART diffractometer equipped with an APEX II CCD area detector operated with monochromated Mo\,$K_{\alpha}$ radiation ($\lambda$[Mo\,$K_{\alpha}$]\,=\,0.71073\,\AA) at 40~kV and 30~mA. Data were collected with frame widths of 1.0$^{\circ}$ in $\omega$ and $\varphi$, and an exposure of 2 s per frame. Data were integrated and corrected for background, Lorentz, and polarization effects by means of the Bruker programs \emph{APEX2} and \emph{XPREP}. Absorption correction was applied using the empirical spherical model within the \emph{SADABS} program~\cite{Sheldrick_2007}. Diffraction data for \textbf{2} were collected at 293~K using a Rigaku Oxford Diffraction Xcalibur diffractometer operated with monochromated Mo\,$K_{\alpha}$ radiation ($\lambda$[Mo\,$K_{\alpha}]$\,=\,0.71073\,\AA) at 50 kV and 40 mA and equipped with an Eos CCD area detector. Data were collected with frame widths of 1.0$^{\circ}$ in $\omega$ and $\varphi$, and an exposure of 2 s per frame. Data were integrated and corrected for background, Lorentz, and polarization effects. An empirical absorption correction based on spherical harmonics implemented in the SCALE3 ABSPACK algorithm was applied in \emph{CrysAlisPro} program~\cite{CrysAlisPro}. The unit cell parameters of \textbf{1} and \textbf{2} (Table~\ref{Table_lat_param}) were determined and refined by least-squares techniques. The structures were solved by direct methods and refined using the \emph{SHELX} program~\cite{Sheldrick_2015} incorporated in the \emph{OLEX2} program package~\cite{Dolomanov_2009}. The final models included coordinates, see Table~\ref{Table2}, and anisotropic displacement parameters for all atoms. Selected interatomic distances are listed in Table~\ref{Table_interatomic_dist}.
It should be noted that in some cases highly redundant XRD data (full sphere and $I/\sigma >$ 30--40) allow refining the site occupancy factors for close, even neighbor, elements from the Periodic Table~\cite{Linden2017}. Supplementary crystallographic data have been deposited in the Inorganic Crystal Structure Database (CSD 1884474 (\textbf{1}) and 1884475 (\textbf{2})) and can be obtained from the Cambridge Crystallographic Data via https://www.ccdc.cam.ac.uk/structures/.

\begin{figure}[t!]
\begin{center}
\includegraphics[width=0.85\columnwidth]{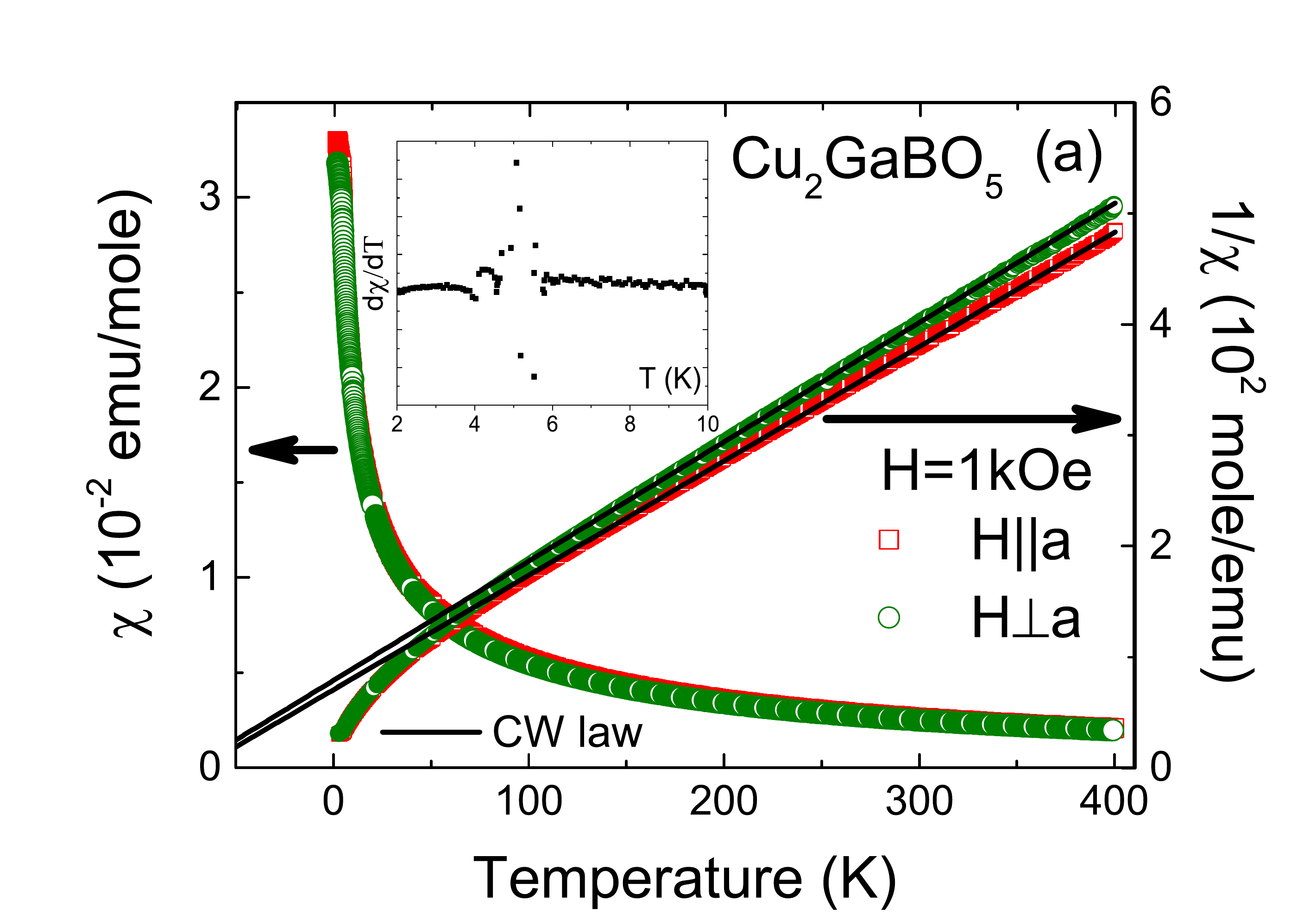}\bigskip\newline
\includegraphics[width=0.85\columnwidth]{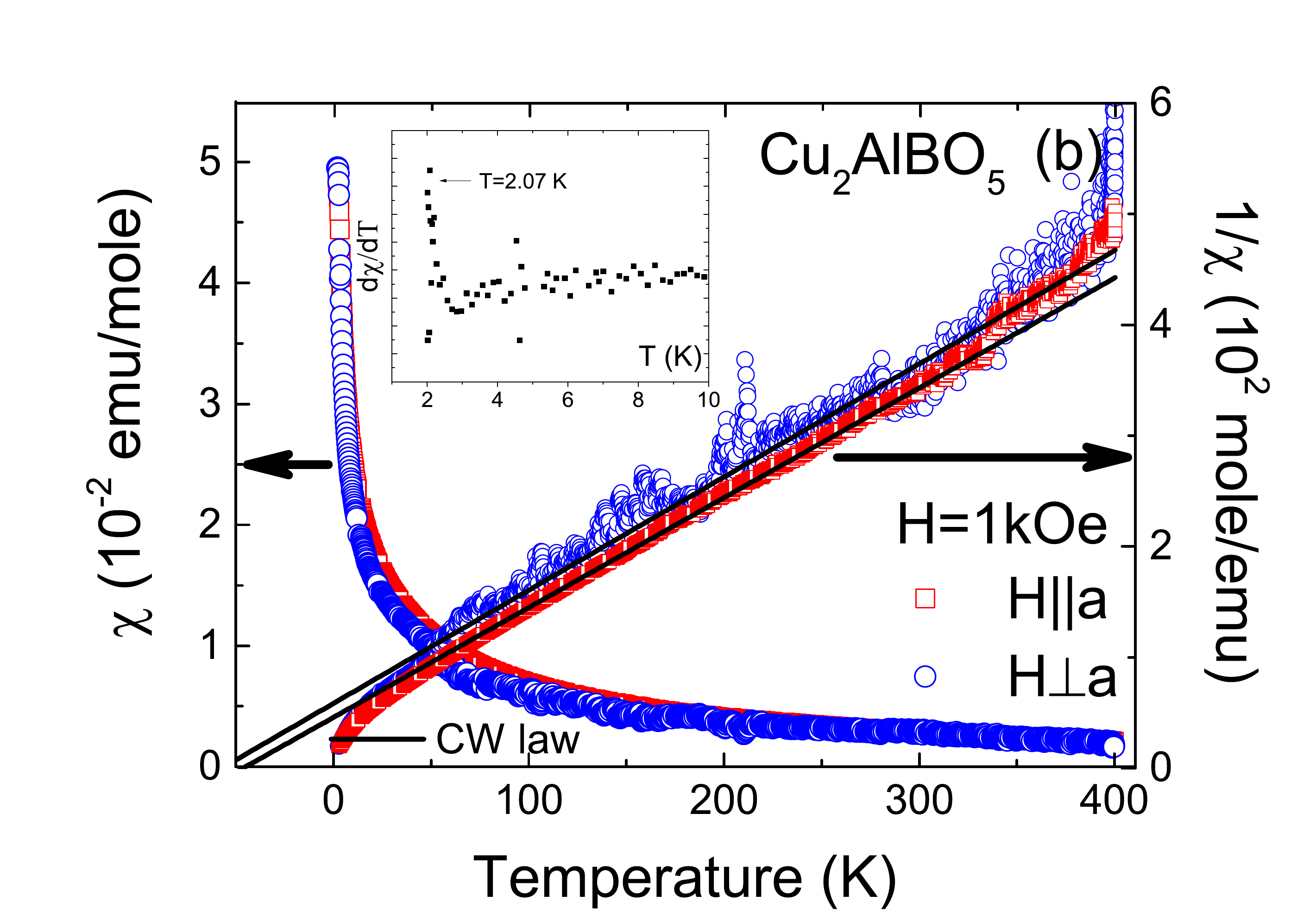}\vspace{-5pt}
\end{center}
\caption{Temperature dependencies of the magnetic susceptibility and inverse magnetic susceptibility in (a) Cu$_2$GaBO$_5$ and (b) Cu$_2$AlBO$_5$ ludwigites measured in FC regime in the magnetic field ($H=1$\,kOe) applied parallel and perpendicular to the crystallographic $a$ axis. The inset shows temperature dependence of the derivative magnetic susceptibility at low temperatures.}
\label{fig1}
\end{figure}

\subsection{Magnetic susceptibility}

\begin{figure}[b!]\vspace{-2pt}
\centerline{\includegraphics[width=0.85\columnwidth]{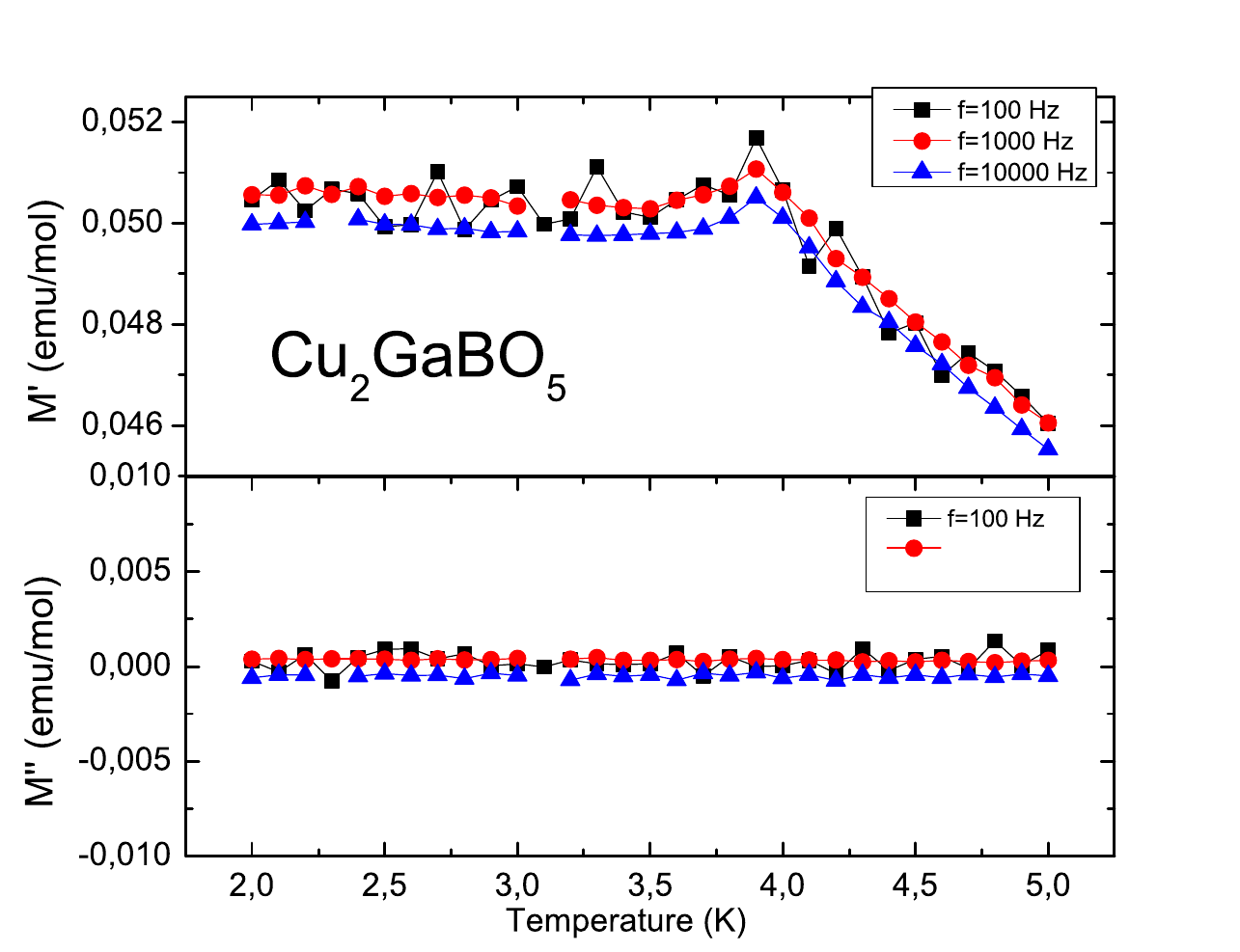}}
\caption{Temperature dependence of the real and imaginary parts of the AC magnetization as a function of frequency at low temperatures for $H=4$~Oe in Cu$_2$GaBO$_5$.}
\label{fig2}
\end{figure}

The magnetization $M$ of single crystalline samples was measured on the commercial Physical Properties Measurements System (PPMS-9 device) within a temperature range 2\,K\,$\leq T \leq$\,400\,K in zero-field-cooled (ZFC) and field-cooled (FC) regimes in magnetic fields $H$ up to 9\,T. Figure~\ref{fig1} shows the FC magnetic susceptibility $\chi$\,=\,$M/H$ of Cu$_2$GaBO$_5$ and Cu$_2$AlBO$_5$ ludwigites as a function of temperature for the magnetic field applied parallel and perpendicular to the crystallographic $a$-axis.
The temperature dependence of the derivative magnetic susceptibility $\partial \chi/ \partial T$\,=\,$\partial M/\partial H$ measured in the FC regime at low temperatures is presented in the inset of Fig.~\ref{fig1}.
The magnetic phase transition temperature was obtained as the susceptibility anomaly which corresponds to the maximum of the derivative magnetic susceptibility: $T=4.1$--5.5~K for Cu$_2$GaBO$_5$, in agreement with our published results~\cite{Eremina_2019}, and $T=2.08$\,K for Cu$_2$AlBO$_5$, respectively (Fig.~\ref{fig1}). The obtained temperature for Cu$_2$GaBO$_5$ is different from the previously observed N\'{e}el temperature $T_{\rm N}=3.4$\,K~\cite{Petrakovskii_2009}. The difference in N\'{e}el temperatures can result from the quality of the sample or from the destruction of the AFM order by a sufficiently weak magnetic field.

\begin{figure}[t!]
\centerline{\includegraphics[width=0.8\columnwidth]{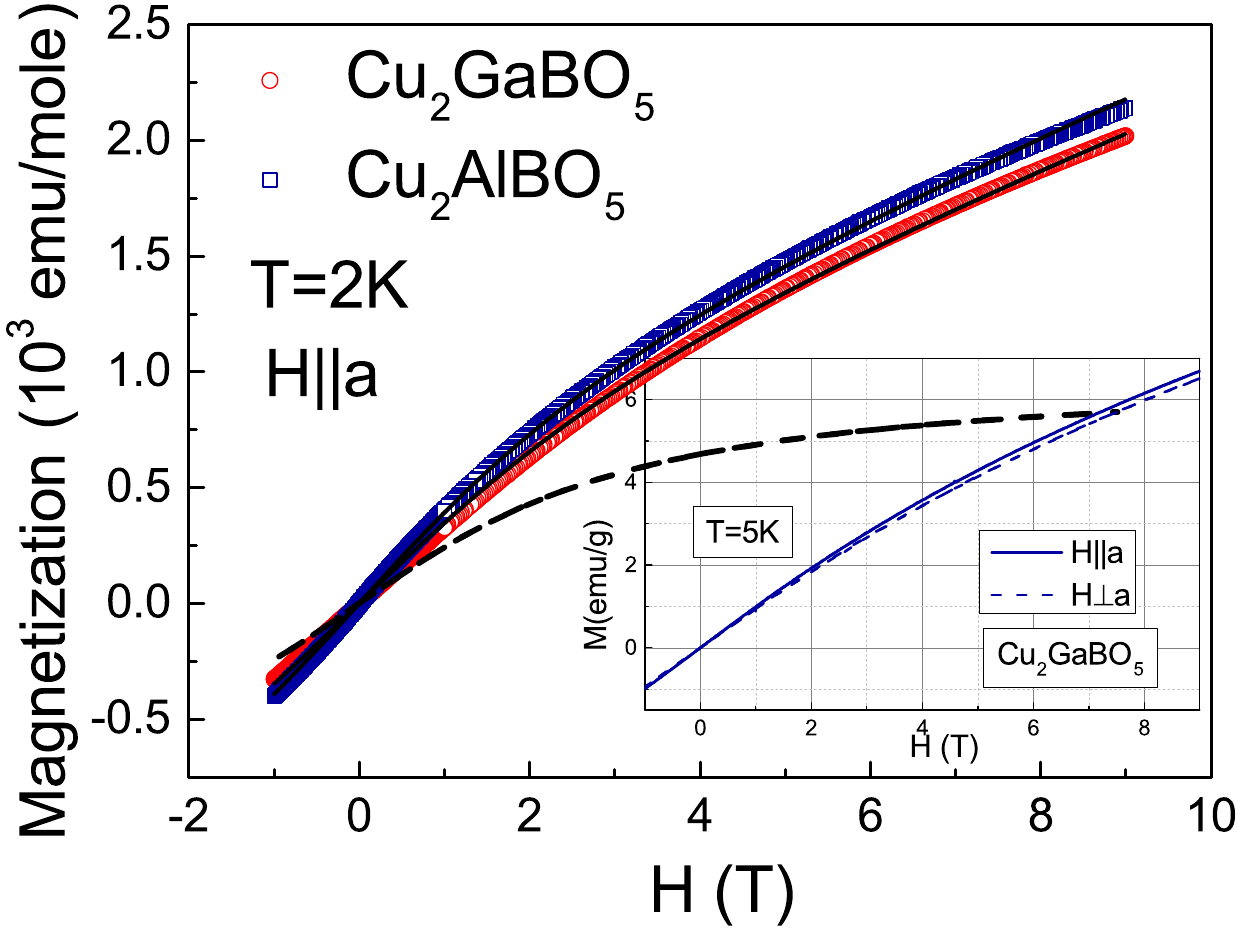}}
\caption{Magnetic field dependence of the magnetization measured at low temperatures. The solid lines are sums of the magnetic and paramagnetic contributions. The dashed line is the magnetic contribution.}
\label{fig3}
\end{figure}

It is possible to unambiguously indicate the type of magnetic phase transition only by studying the temperature dependence of the magnetization on alternating current.
Figure~\ref{fig2} presents the temperature dependence of the AC magnetic susceptibility
measured with $H=4$\,Oe. The presence of a peak in the real part of the magnetization and its absence in the imaginary part of the magnetization indicates the transition from an antiferromagnetic to the paramagnetic state for Cu$_2$GaBO$_5$ single crystal at 4.1\,K. The magnetic field dependence of the magnetization for Cu$_2$GaBO$_5$ and  Cu$_2$AlBO$_5$ is presented~in~Fig.~\ref{fig3}.

\subsection{Specific heat}

\begin{figure}[t!]
\begin{center}\vspace{-4pt}
\includegraphics[width=0.75\columnwidth]{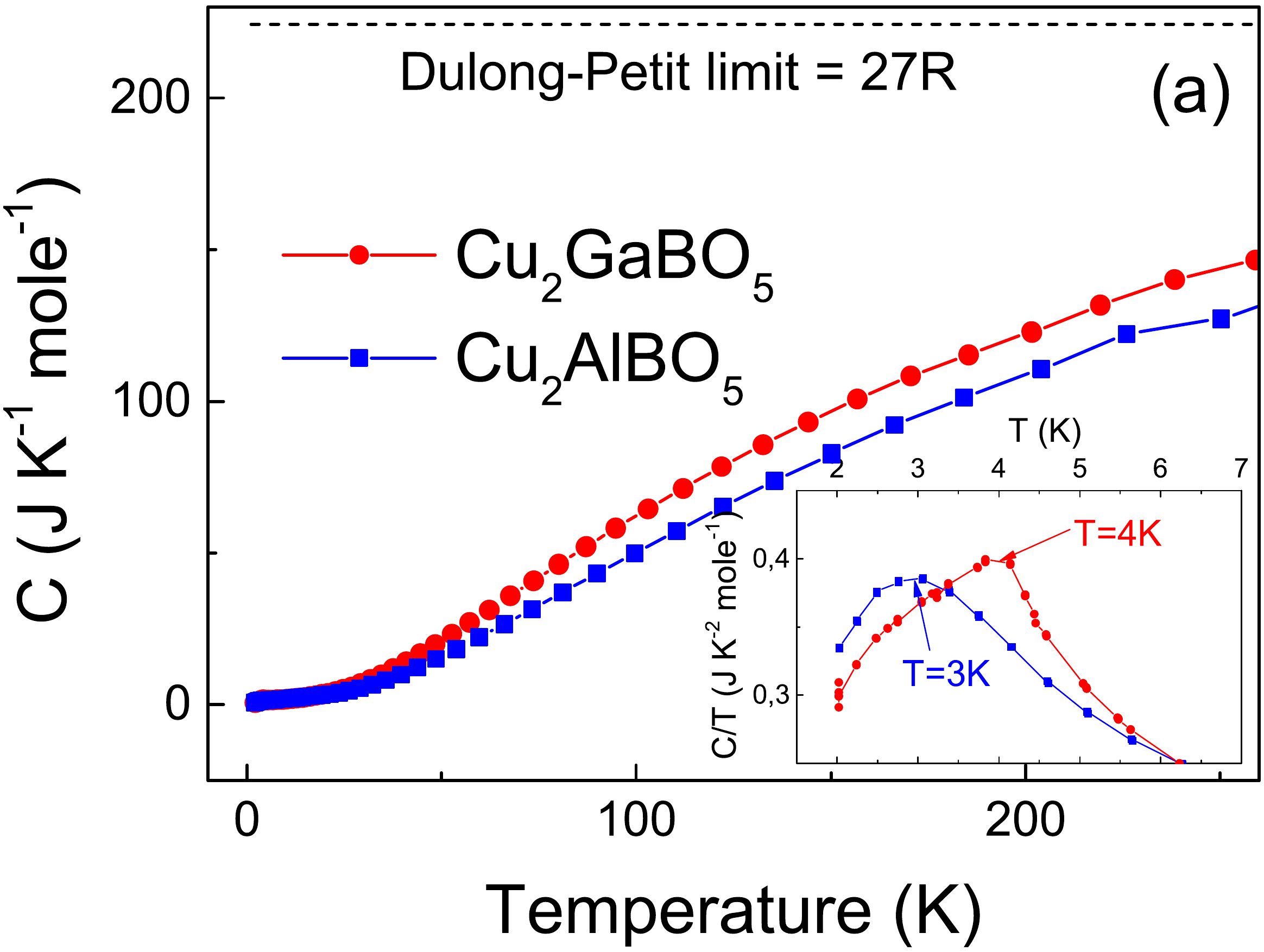}\bigskip\\
\includegraphics[width=0.75\columnwidth]{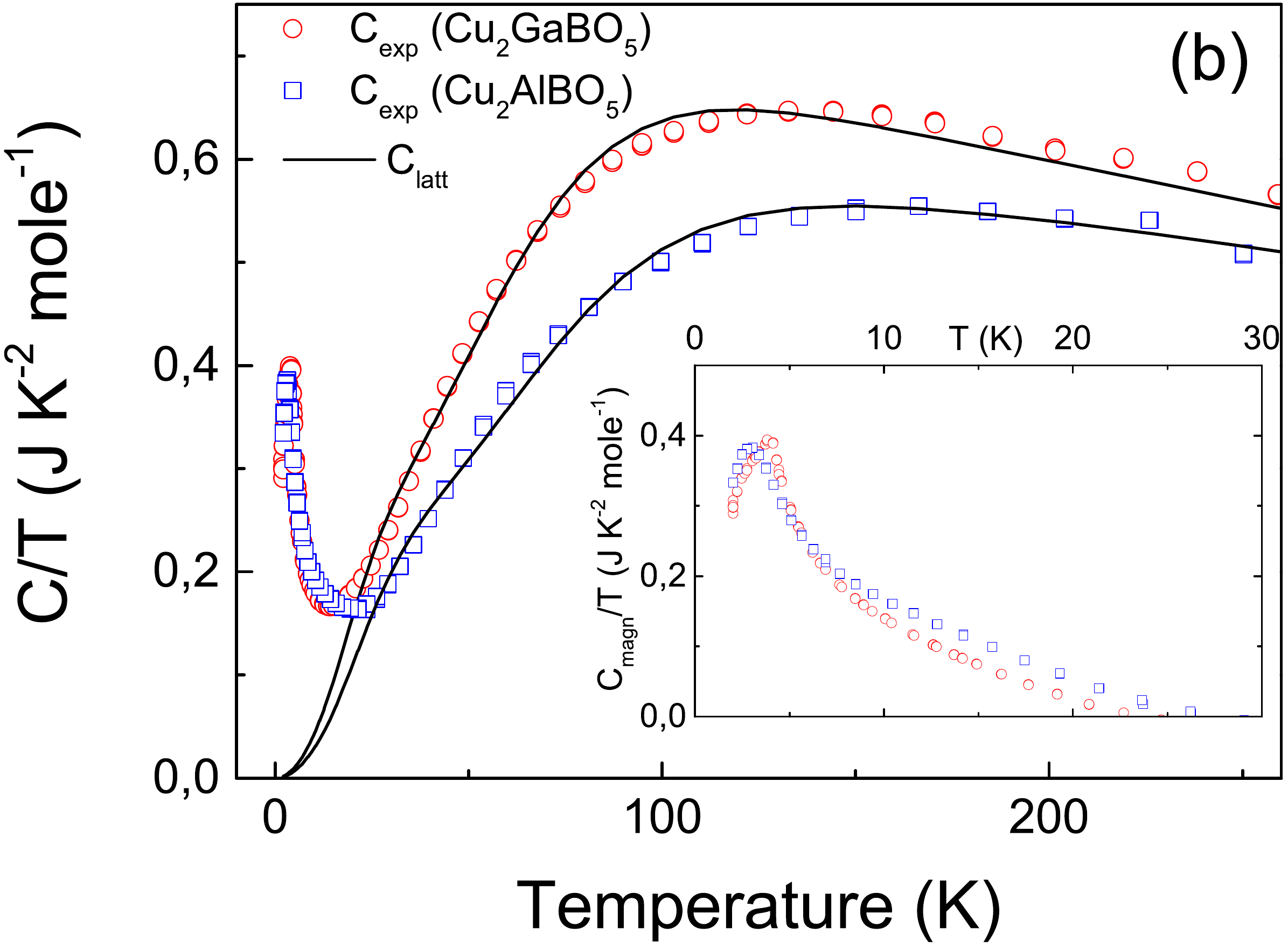}\vspace{-8pt}
\end{center}
\caption{(a)~Temperature dependence of the specific heat $C(T)$ of Cu$_2$GaBO$_5$ and  Cu$_2$AlBO$_5$ measured in zero magnetic field. Inset: Specific heat of Cu$_2$GaBO$_5$ and  Cu$_2$AlBO$_5$ in the temperature range 2\,K\,$\leq$\,$T$\,$\leq$\,10\,K. (b)~Specific heat in representation $C/T$ as a function of $T$, the black solid line is the fitting curve (see details in the text) Inset: low-temperature magnetic specific heat $C_{\rm M}/T$ versus $T$ in zero external magnetic field after subtracting the calculated lattice contribution.}
\label{fig4}
\end{figure}

\begin{figure}[t!]\vspace{-2pt}
\centerline{\includegraphics[width=0.7\columnwidth]{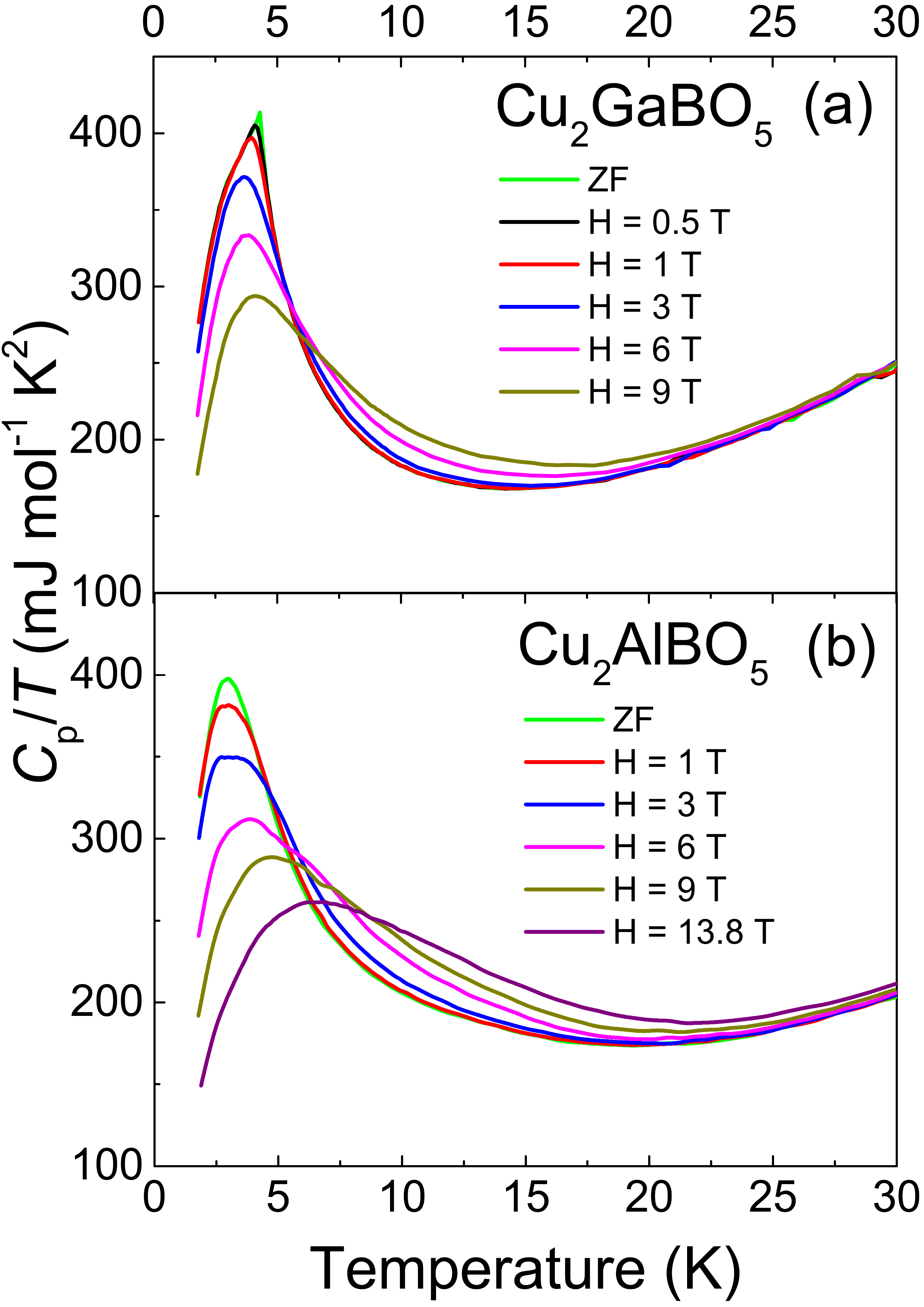}}
\caption{Magnetic contribution to the specific heat $C_{\rm M}(T)/T$ of (a)~Cu$_2$GaBO$_5$ and (b)~Cu$_2$AlBO$_5$ measured both on field cooling (FC) and in zero magnetic field.}
\label{fig5}
\end{figure}

The specific heat was measured by the relaxation method using a Physical Property Measurement System (PPMS) (Quantum Design) in the temperature range 1.8\,K\,$<$\,$T$\,$<$\,400\,K and in magnetic fields up to 13.8\,T. Figure~\ref{fig4} shows the specific heat $C(T)$ as a function of temperature for Cu$_2$GaBO$_5$ and  Cu$_2$AlBO$_5$ ludwigites measured in zero magnetic field. An anomaly was observed in the $C(T)$ data at $T\approx 4$\,K and $T\approx 3$\,K in Cu$_2$GaBO$_5$ and Cu$_2$AlBO$_5$, respectively (inset in Fig.~\ref{fig4}\,a). The obtained values are close to the corresponding anomalies in the $\partial\chi/\partial T$ vs. $T$ curves. In the accessible magnetic-field range, the anomaly is field dependent, indicating the broadening and the decreasing of the peak intensity with increasing of the applied magnetic field (Fig.~\ref{fig5}). This suggests that the small magnetic field (on the order of 2.5~T) destroys of the long-range AFM order.

\section{Discussion}

\begin{figure}[t]
\centerline{\includegraphics[width=0.85\columnwidth]{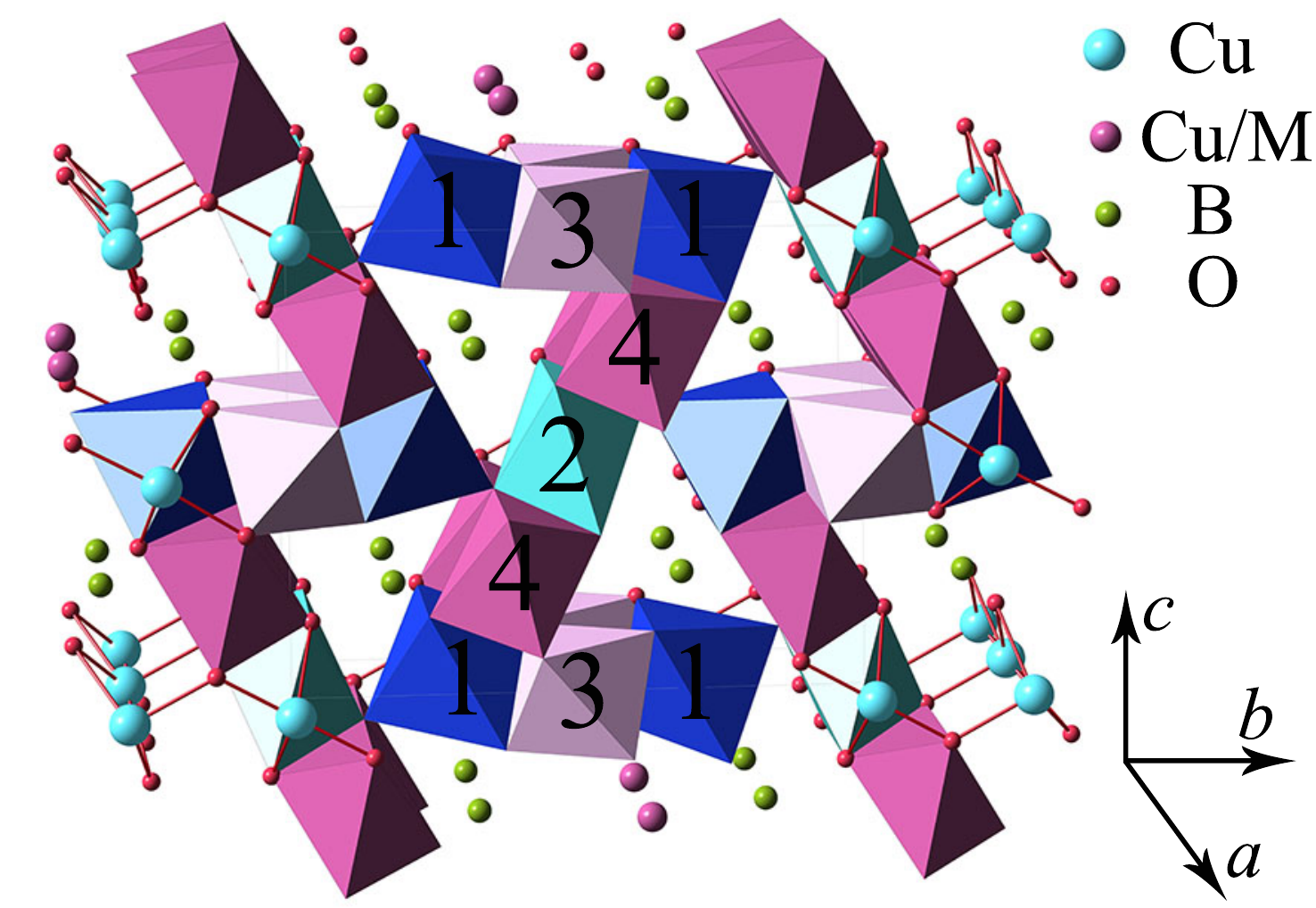}}
\caption{Crystal structure of Cu$_2M$BO$_5$ ($M$\,=\,Ga, Al) ludwigites. Dark blue and light blue octahedrons surround Cu1/$M$ and Cu2/$M$ positions, light magenta and dark magenta octahedrons surround Cu3/$M$ and Cu4/$M$  positions, respectively.} \label{fig6}
\end{figure}

The unit cell of the investigated ludwigites contains $Z=4$ formula units, so the unit cell can contain up to twelve divalent cations (Cu$^{2+}$, $3d^9$) with spin $S = 1/2$. Each magnetic Cu$^{2+}$ ion is surrounded by six oxygen ions forming a strongly distorted octahedron.
We can identify four types of structurally nonequivalent oxygen octahedra, which correspond to four atomic sites of copper ions.
Four types of oxygen octahedra form zigzag walls which are presented in Fig.~\ref{fig6}.
Interatomic distances between cations and anions are given in Table~\ref{Table_interatomic_dist}.
The obtained here crystallographic parameters are close to the previously published parameters for Cu$_2$GaBO$_5$ ludwigites.
In addition to Refs.~\cite{Schaefer_1995} and \cite{Hriljac_1990} we have analyzed the selective distribution of cations on metal sites.
The distinctive features of the structures of Cu$_2$GaBO$_5$ (\textbf{1}) and Cu$_2$AlBO$_5$ (\textbf{2}) are the selective distribution of Cu, Ga and Al cations (Table~\ref{Table2}). M1, M2, M3 and M4 sites in the structure are presented in Fig.~\ref{fig4}. M1 and M2 sites are fully occupied by Cu atoms (site-scattering factors = 28.7 and 28.8 epfu); whereas M3 and M4 sites in the structure of \textbf{1} are predominately occupied by Ga atoms with less amount of Cu (Ga:Cu = 0.66:0.34 and 0.71:0.29, respectively). M1 and M2 sites are predominately occupied by Cu atoms with significantly less amount of Al (Cu:Al = 0.88:0.12 and 0.86:0.14, respectively); whereas M3 and M4 sites in the structure of \textbf{2} are predominately occupied by Al atoms with less amount of Cu (Cu:Al = 0.34:0.66 and 0.33:0.67, respectively).
The presence of  copper-ion chains in the structure \textbf{1} (Cu$_2$GaBO$_5$) can affect the magnetic properties of the compound.

\begin{table}[b!]
\caption{\label{Table_Magnetization} Fitting parameters of the temperature dependence of the magnetic susceptibility for some compounds of the ludwigite family. The magnetic ordering temperature $T_{\rm MO}$ and the fit parameter $\Theta$ are given in K; Curie constants are given in emu$\cdot$K/mol.\smallskip}
\centerline{
\begin{tabular}{l| l r r c c c c c c c c c c c}\hline \hline
                      & $T_{\rm N}$  & $\Theta_{\parallel a}$ &$\Theta_{\perp a}$ &$C_{\parallel a}$  &$C_{\perp a}$  & Ref.             \\ \hline
Cu$_2$GaBO$_5$       &  4.0   &  --69                   &  --74              & 0.97              & 0.93          & this work     \\
Cu$_2$AlBO$_5$       &  2.8   &  --47                   &  --58              & 1.01              & 0.98          & this work     \\ \\
Cu$_2$GaBO$_5$       & 3.4    &  --68                   &  --54              &                   &               & \cite{Petrakovskii_2009} \\
FeMg$_2$BO$_5$       & 8      & --170                   & --170              &                   &               & \cite{Neuendorf_1997} \\
\hline \hline
\end{tabular}}
\end{table}

From magnetic susceptibility measurements (Fig.~\ref{fig1}) we can see that
for both samples $\chi$ can be well fitted by the Curie-Weiss law $\chi=C/(T-\Theta)$ above 50\,K, that is
confirmed by the linear temperature dependence of the inverse magnetic susceptibility (Fig.~\ref{fig1}).
The fitting parameters--Curie constant $C$ and Curie-Weiss temperature $\Theta$ are summarized in Table~\ref{Table_Magnetization}.
The Curie-Weiss temperature $\Theta$ is negative, which indicates that exchange interactions between copper spins are predominantly antiferromagnetic in the investigated samples.
The effective magnetic moment estimated as $\mu_{\rm eff} = \sqrt{3k_{\rm B}C/N_{\rm A}}$ is (2.72$\div$2.78)$\mu_{\rm B}$ and (2.81$\div$2.85)$\mu_{\rm B}$ for
the two copper ions in the formula unit of Cu$_2$GaBO$_5$ and Cu$_2$AlBO$_5$, respectively.
Theoretically, the effective magnetic moment of copper ions can be evaluated as:
\begin{equation}\vspace{-2pt}
\mu_{\rm eff}({\rm Cu}^{2+})=g\sqrt{N_{\rm S}S(S+1)}=2.69~\mu_{\rm B}.
\end{equation}\enlargethispage{3pt}
Since only the copper spins contribute to the magnetization, we use the value $g=2.2$ which is characteristic for copper ions in the octahedron environment formed by oxygen ions and $S=1/2$ for Cu$^{2+}$, $N_{S}$=2
is the number of ions with spin $S$ in the chemical formula unit.
We can see that the experimentally obtained values of the effective magnetic moments are close to the theoretically predicted ones.
As there are 4 inequivalent Cu sites, all of them could have different $g$-factor values and different magnetic moments.
It should be perhaps made clear that these values refer to some average moments.
The effective magnetic moments for Cu$_2$AlBO$_5$ is higher than for Cu$_2$GaBO$_5$.
At lowest temperatures, we can also see that the susceptibility $\chi$ in Cu$_2$AlBO$_5$ approaches a value of 5.58$\times$10$^{-2}$~emu/mol, that is 1.5 times higher than for  Cu$_2$GaBO$_5$ (Fig.~\ref{fig1}).
This indicates a more significant paramagnetic contribution from the random distribution of copper ions and defects in the samples. The number of defects was changed from sample to sample.

The dependencies of magnetization on the magnetic field were measured on Cu$_2$GaBO$_5$ and  Cu$_2$AlBO$_5$, and it was shown that the paramagnetic component for Cu$_2$AlBO$_5$ was slightly larger than in Cu$_2$GaBO$_5$.
Magnetic field dependencies of the magnetization in these compound at temperature  2\,K, are presented in Fig.~\ref{fig3}. At 2\,K the magnetizations
have been described as a sum of two contributions (see Fig.~\ref{fig3}) by the formula $B = B_{\rm m}+B_{\rm pm}$. There $B_{\rm m} = \frac{2B_{\rm S}}{\pi}\tan^{-1}\left[(H\pm H_{\rm C})/H_T\right]$ is the magnetic contribution~\cite{Geiler_2006}, with $B_{\rm S}$ is the saturation magnetization and $H_T$ is the inner local field of uniaxial anisotropy, $H_{\rm C}$ is the coercive field and $B_{\rm pm} = \chi H$ is the paramagnetic contribution from defects. Fitting parameters are equal to $B_{\rm S}=0.1$\,T, $B_{\rm S}=0.095$\,T; $H_{T}=2.5$\,T, $H_{T}=2.8$\,T; $\chi=1.5$~emu/(K$\cdot$mol), $\chi=1.4$~emu/(K$\cdot$mol); and $H_{\rm C}=0$ for Cu$_2$GaBO$_5$ and Cu$_2$AlBO$_5$, respectively. As follows from our description, the weak external magnetic field 2.5\,T destroys the long magnetic order and when the magnetic field is increased more, the spins have been polarized along the external magnetic field.

In the Cu$_2$GaBO$_5$ compound gallium ions are nonmagnetic and only magnetic moments of copper ions are ordered with $T_{\rm N}=4$\,K. Temperature dependencies of image and real magnetization
for this compound are displayed in Fig.~\ref{fig2}. As expected for the AFM structure, the peak is absent in the imaginary part \cite{Balanda_2013}.

As we know from X-ray diffraction measurements, there are linear chains of Cu$^{2+}$ ions in the M1 and M2 sites of the investigated sample, located along the crystallographic \emph{a}-axis (Fig.~\ref{fig6}).
Typically for 1D antiferromagnetic Heisenberg chains of localized spins the magnetic susceptibility exhibits a maximum at a temperature comparable to the intrachain exchange, as it was in the case of the copper-based compounds CuTe$_2$O$_5$~\cite{Deisenhofer_2006}, CuTa$_2$O$_6$~\cite{Golubev_2017}, Na$_2$Cu$_2$TeO$_6$~\cite{Xu_2005}, Na$_3$Cu$_2$SbO$_6$~\cite{Schmitt_2014}, where the temperature dependence of the susceptibility could be approximated using the model of an AFM spin $S$\,=\,1/2 chain~\cite{Johnston_2000}. As we can see from Fig.~\ref{fig1}, the magnetic susceptibility of Cu$_2$GaBO$_5$ and  Cu$_2$AlBO$_5$ does not exhibit a broad maximum but displays a sharp cusp at 4.0 and 2.8~K, respectively.\enlargethispage{3pt}

Also in contrast to the investigated here bimetallic monomagnetic ludwigites with the temperature of the AFM ordering $T_{\rm N}=2$--4~K and previously investigated FeMg$_2$BO$_5$ with the N\'{e}el temperature $T_{\rm N}\simeq8$\,K~\cite{Neuendorf_1997} the magnetic phase transition in bimagnetic ludwigites was observed at much higher temperatures: 92\,K -- for Cu$_2$MnBO$_5$~\cite{Moshkina_2017},
81$-$92\,K -- for Mn$_{3-x}$Ni$_x$BO$_5$~\cite{Bezmaternyk_2014}. For iron-containing bimagnetic ludwigites ($M_2$FeBO$_5$, $M$\,=\,Ni, Cu, Co) it was observed that the ordering temperatures of the Fe$^{3+}$ sub-system in each compound is higher than that of the respective $M$ sub-system~\cite{Continentino_1999, Fernandes_2000,Fernandes_1998, Freitas_2009}. The magnetic phase transition temperature in homometallic magnetic ludwigites is also rather high: 42\,K -- for  Co$_3$BO$_5$~\cite{Freitas_2008} and 112\,K -- the temperature of the partial magnetic transition for Fe$_3$BO$_5$, while the whole system of Fe moments
become magnetically ordered at $T_{\rm N}=70$\,K~\cite{Guimaraes_1999,Freitas_2008}.

The investigations of thermodynamic properties showed that at 300\,K the specific heat is still considerably lower than the expected high-temperature value $3Rs\,=\,225$\,J/(mol$\cdot$K) for the phonon contribution given by the Dulong-Petit law (Fig.~\ref{fig4}), indicating contributions to the phonon-density of states from higher-lying lattice modes. Here, $R$ denotes the gas constant and $s$ the number of atoms per formula unit.
We assume that the total heat capacity originates from two different contributions, a lattice contribution $C_{\rm latt}$ due to acoustic and optical phonons and a magnetic contribution $C_{\rm M}$ corresponding to the thermal population of excited magnetic states. We expect that the magnetic contribution is small compared to the lattice contribution as it was in the case of other copper-based low-dimensional systems~\cite{Eremina_2011,Lysogorskiy_2014}. The straightforward method to unambiguously extract the magnetic contribution from the experimental data is difficult to realize because a specific heat data for non-magnetic reference material (Mg$_2$AlBO$_5$~\cite{Bloise_2010} or Zn$_2$AlBO$_5$) is not available. The lattice contribution C$_{\rm latt}$ was approximated following standard procedures~\cite{Gopal_1966} with a minimized set of fit parameters only using a sum of one isotropic Debye ($C_{\rm D}$) accounting for the 3 acoustic phonon branches and two isotropic Einstein terms ($C_{\rm E1}$, $C_{\rm E2}$) averaging the $3s-3\,=\,24$ optical phonon branches:
\begin{eqnarray}
\label{math_spheat}
C\,=\,C_{\rm latt}+ C_{\rm M}, \nonumber\\
C_{\rm latt}\,=\,\alpha_{\rm D}\cdot C_{\rm D}\,+\,\alpha_{\rm E1}\cdot C_{\rm E1}\,+\,\alpha_{\rm E2}\cdot C_{\rm E2}.
\end{eqnarray}

For further reducing the number of free fit parameters, the ratio between these terms was fixed to $\alpha_{\rm D}$\,:\,$\alpha_{\rm E1}$\,:\,$\alpha_{\rm E2}$ = 1\,:\,4\,:4 to account for the 3$s$\,=\,27 degrees of freedom per formula unit. For $s$\,=\,9 atoms formula unit, the ratio between acoustical (Debye) and optical (Einstein) contributions is naturally fixed as 1\,:\,8. The weight distribution between the optical contributions is chosen in such a way that the degrees of freedom have been equally distributed between the higher Einstein modes. The resulting fit curve (solid line in Fig.~\ref{fig3}\,b) describes the data satisfactorily. For the respective Debye and Einstein temperatures we obtained $\Theta_{\rm D}=166.3$\,K, $\Theta_{\rm E1}=338.4$\,K, $\Theta_{\rm E2}=1009.3$\,K -- for Cu$_2$GaBO$_5$ and $\Theta_{\rm D}=189.5$\,K, $\Theta_{\rm E1}\,=\,401.3$\,K, $\Theta_{\rm E2}=1108.2$\,K -- for Cu$_2$AlBO$_5$. As one can see, the existence of high-frequency modes at 1009.3\,K or 1108.2\,K agrees well with the fact that the Dulong-Petit value is approached only far above room temperature.

The magnetic contribution to the specific heat $C_{\rm M}$ was obtained as the difference between the experimentally measured data and the calculated by the Eq.~\ref{math_spheat} lattice contribution $C_{\rm M}=C_{\rm exp}-C_{\rm latt}$. The inset in Fig.~\ref{fig4}\,b shows the obtained a such a way the magnetic contribution for zero magnetic field. The temperature dependence of $C_{\rm M}$ for different values of the applied magnetic field is given in Fig.~\ref{fig5}. The magnetic contribution in Cu$_2$AlBO$_5$ has the broad maximum, its width increases with increasing of the magnetic field (Fig.~\ref{fig6}\,a).
We suggest that such a behavior of the specific heat together with the sharp peak in $\chi-T$ curve (Fig.~\ref{fig1}\,b) and the random distribution of copper ions in the crystal structure is due to the AFM transition in Cu$_2$AlBO$_5$ at $T_{\rm N}\approx3$\,K. Previously, the shift of the broad maximum in the temperature dependence of the magnetic specific heat together with the anomaly in $\chi-T$ curve were observed \cite{Binder_1986,Grinenko_2013,Skornia_2018}.

For Cu$_2$GaBO$_5$ we suggest the presence of one type of the extended phase transition. The transition begins to be observed at $T=4.0$\,K in the AFM long-range oder and the sample is completely ordered to the temperature $T_{\rm N}\approx3$\,K,  we attribute this to the random distribution of copper and gallium ions on M3 and M4 positions (see Fig.~\ref{fig6}). The long-range magnetic order is easily destroyed by a magnetic field larger than 2.5\,T, but short-range order regions are preserved. For Cu$_2$AlBO$_5$ we suggest the presence of one type of AFM phase transition. Probably, this is the ordering of AFM clusters formed near a nonmagnetic impurity (aluminum). Similar behavior was observed in quasi-one-dimensional magnetic CuGeO$_3$ with impurity and defects~\cite{Demishev_2009,Glazkov_1998}.

\section{Summary}

Here we presented the investigations of single crystals of Cu$_2$GaBO$_5$ and Cu$_2$AlBO$_5$ oxyborates with the ludwigite structure synthesized by the flux technique. The distinctive features of the investigated structures are the selective distribution of Cu, Ga and Al cations. The unit cell of Cu$_2$GaBO$_5$ and Cu$_2$AlBO$_5$ contains four nonequivalent crystallographic sites of metal ions. Two sites in the structure of Cu$_2$GaBO$_5$ are predominantly occupied by Ga atoms with less amount of Cu (Ga:Cu = 0.71:0.29 and 0.66:0.34, respectively); whereas other sites are fully occupied by Cu atoms. For Cu$_2$AlBO$_5$ all sites are partially  occupied by Al and Cu atoms. M1 and M2 sites are predominately occupied by Cu atoms with significantly less amount of Al (Cu:Al = 0.88:0.12 and 0.86:0.14, respectively); whereas M3 and M4 sites are predominantly occupied by Al atoms with less amount of Cu (Cu:Al = 0.34:0.66 and 0.33:0.67, respectively). The magnetic properties of the investigated homomagnetic copper ludwigites are discussed in comparison with known heterometallic bimagnetic ludwigites.

The magnetic measurements showed that the effective magnetic moment and low-temperature magnetic susceptibility for Cu$_2$AlBO$_5$ is higher than for Cu$_2$GaBO$_5$. This is probably due to changes in the g-factors of copper ions surrounded by more distorted octahedron from oxygen ions.

The analysis of the phonon contribution to the specific heat was performed, that allowed to separate the magnetic contribution to the specific heat for both compounds. The Debye and Einstein temperatures were obtained from the analysis of the temperature dependence of the specific heat. The joint analysis of low-temperature data on magnetic susceptibility and magnetic contribution to the specific heat showed that antiferromagnetic clusters which formed near defects in Cu$_2$AlBO$_5$ and Cu$_2$GaBO$_5$ go into a antiferromagnetic state at $T_{\rm N}\approx3$\,K.  The magnetic phase transitions was started in Cu$_2$GaBO$_5$ at $T=4.0$\,K, which can be the transition to the antiferromagnetically ordered state in quasi one-dimensional chain formed by copper ion along the $a$-axis. An external magnetic field above 2.5\,T apparently destroys the long-range antiferromagnetic order, but short-range magnetic order is preserved.

\section{Acknowledgments}

The reported study was supported by the Russian Foundation for Basic Research (RFBR), grant No 17-02-00953. T.P.G., R.M.E. acknowledge the financial support from the government assignment for FRC Kazan Scientific Center of RAS. D.S.I. acknowledges funding from the German Research Foundation through the Collaborative Research Center SFB 1143 in Dresden (project C03) and the W\"urzburg-Dresden Cluster of Excellence on Complexity and Topology in Quantum Matter -- \textit{ct.qmat} (EXC 2147, project-id 39085490). The XRD and EDX measurements have been performed at the X-ray Diffraction Centre and Centre for Microscopy and Microanalysis of the St. Petersburg State University. The magnetometer studies were performed by the subsidy allocated to Kazan Federal University for the state assignment in the sphere of scientific activities.

\end{document}